# High-precision laser spectrum analyzer via digital decoherence

ZHONGWANG PANG,[1,2] CHUNYI LI,[1,2] HONGFEI DAI,[1,2] WENLIN LI,[1,2] DONGQI SONG,[1,2] FEI MENG,[3] YIGE LIN,[3] AND BO WANG[1,2,*]

[1]State Key Laboratory of Precision Space-time Information Sensing Technology, Department of Precision Instrument, Tsinghua University, Beijing, China, 100084
[2]Key Laboratory of Photonic Control Technology (Tsinghua University), Ministry of Education, Beijing, China, 100084
[3]Division of Time and Frequency Metrology, National Institute of Metrology, Beijing, China, 100029
*bo.wang@tsinghua.edu.cn

**Abstract:** With the continuous advancement of laser technology, accurately evaluating the noise spectrum of high-performance lasers has become increasingly challenging. In this work, we demonstrate a high-precision laser spectrum analyzer based on the proposed digital decoherence method, which can precisely measure the frequency noise spectrum of sub-Hz linewidth lasers. In addition, it has broad wavelength compatibility, which enables convenient switching between lasers with different center wavelengths. Its performance is validated through measurements of ultra-stable lasers. Based on the measured frequency noise power spectral density, a β-line linewidth is determined to be 570 mHz at 10-second observation time, and the minimum observable linewidth is calculated to be 133 mHz. The system's noise floor is evaluated to be 210 mHz β-line linewidth at 25-second observation time, and a minimum observable linewidth of 39 mHz.

## 1. Introduction

Since their invention, lasers have significantly impacted modern science and industrial processes, finding widespread applications in areas such as metrology [1-3], communications [4], and sensing [5,6]. Over the past few decades, rapid advancements in frequency stabilization techniques have further enhanced the performance of lasers [7–12], making it increasingly challenging to accurately analyze the laser spectrum. Typically, the phase or frequency noise power spectral density (PSD) can reveal the full spectral characteristics of a laser, and is a key metric for evaluating lasers. Besides, the linewidth derived from the noise PSD is also employed as a convenient but rough indicator of laser quality. Due to the fundamental limitations in direct optical frequency detection, these key metrics are typically measured by interfering two laser beams.

One common approach to measure a laser's noise PSD is direct comparison with one or more reference lasers. In heterodyne detection, the reference laser must have the stability equal to or better than that of the laser under test (LUT) [13-15]. The three-cornered hat (TCH) method extends this comparison to three lasers, and can evaluate their noise PSD, respectively [16-19]. These two methods (hereafter referred to as the heterodyne methods) provide precise measurements and are widely used for characterizing lasers with sub-Hz linewidths [20]. However, heterodyne methods face a common challenge: the center wavelengths of the lasers under comparison must be closely matched to generate a detectable beat signal. When this condition is unavailable, an optical frequency comb locked to an ultra-stable laser is often used [21, 22]. However, this approach is intricate and costly [23].

In practical applications, the delayed self-heterodyne interferometry (DSHI) method is often used as a low-precision technique for measuring lasers with linewidths above several hundred hertz, primarily due to its wavelength compatibility [24-26]. For DSHI method, the LUT is compared with its own delayed beam, which eliminates the need for additional reference lasers, and avoids the wavelength matching requirement in heterodyne methods. However, DSHI method faces two major challenges when applied to high-performance lasers. The first is the physical difficulty of implementing the necessary delay fiber for laser decoherence, which always spans hundreds or even thousands of kilometers [27]. The second is the attenuation and environmental noise induced by such a long link [28]. Despite various proposed improvements and algorithms [29-32], these inherent limitations still prevent DSHI method from accurately measuring the full noise PSD of lasers with linewidths below 100 Hz.

If the laser's decoherence process can be implemented digitally, it would overcome the limitations caused by the long fiber link, thereby extending the applicability of the DSHI method to ultra-stable lasers with sub-Hz linewidths. In this paper, a digital decoherence concept is proposed. It allows the decoherence process to be realized not only through physical delay but also by digital delay, which simplifies the realization of decoherence, especially for long-coherence systems. The proposed method preserves the key advantages of DSHI, such as system simplicity and broad wavelength compatibility, thereby enabling convenient switching between lasers with different center wavelengths. At the same time, the measurement precision is significantly improved, reaching the level of heterodyne methods and enabling full noise PSD characterization of lasers with sub-Hz linewidths. A series of experiments are carried out to validate the effectiveness of the proposed method.

First, to verify the measurement correctness of the proposed method, we compare it with TCH method. Three lasers with different center wavelengths from NKT Photonics (Koheras BASIK X15) [33] are measured. The frequency noise PSD results are consistent between the two methods. Second, we demonstrate the capability of the proposed digital decoherence method through measuring the ultra-stable laser designed for optical clock at the National Institute of Metrology of China (NIM). The measured β-line linewidth is 570 mHz at an observation time of 10 s, and the minimum observable linewidth is calculated to be 133 mHz. Noise floor analysis further reveals that the system's measurement capability can reach a β-line linewidth of 210 mHz at 25-second observation time, and a minimum observable linewidth of 39 mHz. Compared to a commercial laser spectrum analyzer [34], the proposed laser spectrum analyzer shows a 40-70 dB reduction in the noise floor.

## 2. Theory

For a DSHI system, as shown in Fig. 1(a), the incoming laser beam is split into a reference beam and a delayed beam. The delayed beam travels through an optical fiber link, experiencing a time delay of $\tau_0$, and then interferes with the reference beam at a photodetector (PD). In this way, the phase characteristics of the LUT at times $t$ and $t-\tau_0$ can be compared and recorded, as represented by $\Delta\varphi_{01}(t) = \varphi(t) - \varphi(t-\tau_0)$. These recorded phase characteristics can be visually displayed as sequential frames of a film in Fig. 1(b), and the film starts at $t_0$, with a duration of $\tau_0$ for each frame. It can be seen that the phase characteristic at $t_0$ can only be compared with that at $t_0 - \tau_0$, and the phase characteristic at $t_0 + (m+1)\tau_0$ can only be compared with that at $t_0 + m\tau_0$. In other words, only phase characteristics separated by $\tau_0$ can be compared, making it difficult to compare phase characteristics with other time intervals. Consequently, to accurately measure the laser's noise PSD, the delay time $\tau_0$ should exceed the coherence time of the LUT. However, high-quality lasers typically have very long coherence times, making it challenging to meet this requirement solely through the physical delay induced by long fibers.

In fact, as shown in Fig. 1(b), the phase characteristics of the laser at all moments have already been recorded. If a comparison can be made between the phase characteristic at $t_0$ in frame 1 and that at $t_0 + m\tau_0$ in frame m+1, this issue will be resolved. This comparison can be achieved through a digital decoherence process. As illustrated in Fig. 2, the film records a set of phase noise difference, with $\Delta\varphi_{ij}(t)$ on the left representing the phase noise difference

$\varphi(t-i\tau_0)-\varphi(t-j\tau_0)$. The orange films record the phase noise differences corresponding to different decoherence times. The black films show the data used in the digital decoherence process, which are identical to the orange films directly above them, except for a temporal shift.

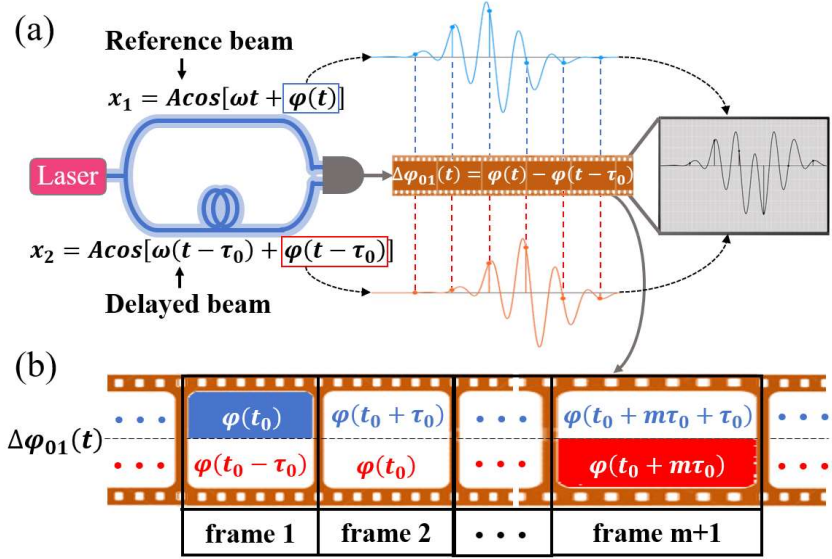

FIG. 1 (a) DSHI system used to obtain the phase noise difference between a laser beam and its delayed beam. The phase fluctuation of the reference beam is shown as the upper blue line, and the phase fluctuation of the delayed beam is shown as the lower red line. The phase fluctuation of the beat signal, i.e., the phase noise difference, is recorded and displayed on the right gray board. (b) We visually use a film to display the recorded phase differences of the two beams at any given moment.

First, by shifting the initial phase noise difference $\Delta\varphi_{01}(t)$ with $\tau_0$ to form the data $\Delta\varphi_{12}(t)$, the phase characteristics at $t_0$, $t_0-\tau_0$ and $t_0-2\tau_0$ can be compared to obtain a phase noise difference $\Delta\varphi_{02}(t)$, which corresponds to the decoherence time of $2\tau_0$, as shown in the second orange film in Fig. 2. Similarly, by shifting the obtained phase noise difference $\Delta\varphi_{02}(t)$ with $2\tau_0$ to form the data $\Delta\varphi_{24}(t)$, and comparing $\Delta\varphi_{02}(t)$ and $\Delta\varphi_{24}(t)$, the phase characteristics at $t_0$, $t_0-2\tau_0$ and $t_0-4\tau_0$ can be compared to obtain a phase noise difference $\Delta\varphi_{04}(t)$, which corresponds to the decoherence time of $4\tau_0$, as shown in the third orange film in Fig. 2. Repeating this process m times will form a phase noise difference $\Delta\varphi_{02^m}(t)$, which corresponds to the decoherence time of $2^m\tau_0$, as shown in the last orange film in Fig. 2. In this way, digital decoherence of the initial phase noise difference is achieved.

The steps of digital decoherence method are summarized as follows:

0. Obtain the initial phase noise difference $\Delta\varphi_{01}(t)$ between two beams via an interference system with a short delay $\tau_0$:

$$\Delta\varphi_{01}(t)=\varphi(t)-\varphi(t-\tau_0). \tag{1}$$

1. Shift the initial phase noise difference $\Delta\varphi_{01}(t)$ by a time delay $\tau_0$ to form the data $\Delta\varphi_{12}(t)$, and compare the data with the initial phase noise difference. This step aims to eliminate the phase fluctuation term $\varphi(t-\tau_0)$ and obtain the phase noise difference with the delay of $2\tau_0$:

$$\begin{aligned}\Delta\varphi_{02}(t) &= \Delta\varphi_{01}(t)+\Delta\varphi_{12}(t) \\ &= \varphi(t)-\varphi(t-\tau_0)+\varphi(t-\tau_0)-\varphi(t-2\tau_0) \\ &= \varphi(t)-\varphi(t-2\tau_0)\end{aligned} \quad (2)$$

2. Repeat step 1. For the m-th step ($m \geq 1$), shift the phase noise difference obtained from the previous step $\Delta\varphi_{02^{m-1}}(t)$ by a time delay $2^{m-1}\tau_0$ to form the data $\Delta\varphi_{2^{m-1}2^m}(t)$, and compare the data with the phase noise difference $\Delta\varphi_{02^{m-1}}(t)$. This forms the phase noise difference with the delay of $2^m\tau_0$:

$$\begin{aligned}\Delta\varphi_{02^m}(t) &= \Delta\varphi_{02^{m-1}}(t)+\Delta\varphi_{2^{m-1}2^m}(t) \\ &= \varphi(t)-\varphi(t-2^m\tau_0)\end{aligned}. \quad (3)$$

In this way, the digital decoherence method can effectively multiply the short fiber-induced physical delay. When $2^m\tau_0$ is longer than the laser's coherence time $\tau_c$, the phase noise PSD of the laser can be evaluated.

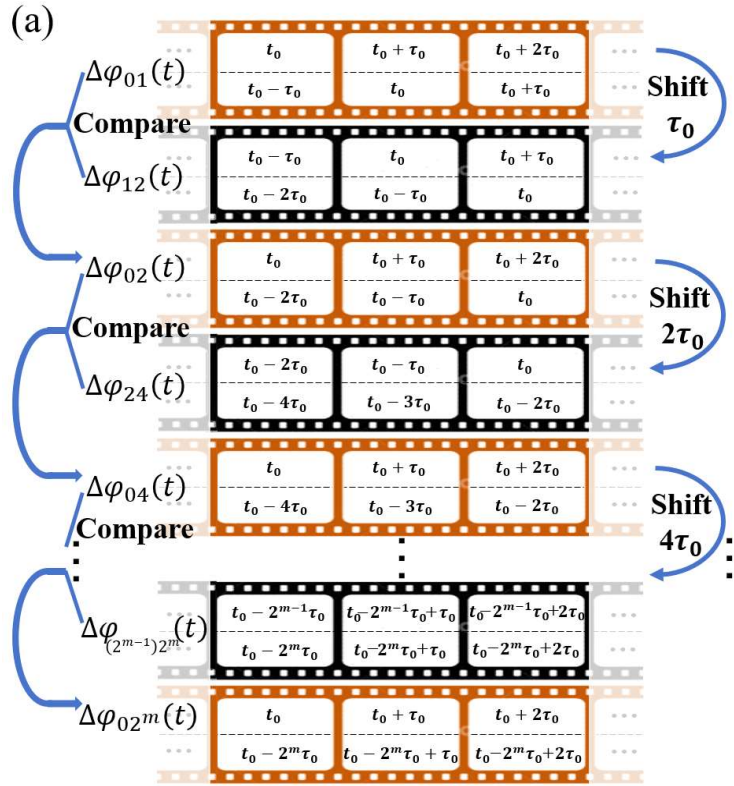

FIG. 2 The principle of digital decoherence method. The time recorded in each frame corresponds to the phase characteristic of the laser at that specific moment, for example, $t_0$ represents $\varphi(t_0)$. The orange films record the phase noise differences corresponding to different decoherence times. The black films are obtained by applying a $\tau_0$ temporal shift to the orange

films directly above them. By comparing the phase characteristics recorded in two films, the same term can be eliminated and the digital decoherence can be achieved.

## 3. Implementation

In practice, the noise from the optical fiber link cannot be ignored. Taking the influence of the fiber link into account, the delayed beam will become $x_2(t) = A\cos[\omega_c(t-\tau_0-\delta\tau_0)+\varphi(t-\tau_0-\delta\tau_0)]$, and the phase fluctuation of the beat signal obtained by the interference system will be:

$$\begin{aligned}\Delta\varphi_{01}(t) &= \varphi(t)-\varphi[t-(\tau_0+\delta\tau_0)]+\omega_c\delta\tau_0 \\ &= \varphi(t)-\varphi(t-\tau_0)+\varphi(t-\tau_0)-\varphi[t-(\tau_0+\delta\tau_0)]+\omega_c\delta\tau_0\end{aligned}, \quad (4)$$

where $\omega_c$ is the angular frequency of the laser, $\tau_0$ is the time delay used during the digital decoherence process, $\delta\tau_0$ is the time delay fluctuation of the fiber link. The consideration of the time delay fluctuation leads to two issues. First, the initial phase noise difference is no longer $\varphi(t)-\varphi(t-\tau_0)$ but $\varphi(t)-\varphi[t-(\tau_0+\delta\tau_0)]$. Second, an error term $\omega_c\delta\tau_0$ appears in the results. For the first issue, it does not affect the effectiveness of the digital decoherence method. To explain this, we decompose $\varphi(t)-\varphi[t-(\tau_0+\delta\tau_0)]$ as shown in Eq. (4). The result consists of two terms: the first term, $\varphi(t)-\varphi(t-\tau_0)$, is the valid signal component used for ideal digital decoherence. The second term, $\varphi(t-\tau_0)-\varphi[t-(\tau_0+\delta\tau_0)]$, represents the error term introduced by the fluctuating time delay of the fiber link. This error term is equivalent to compare the phase characteristics of the LUT at the interval of $\delta\tau_0$. For an optical fiber link with the length of 25 km, given that the effective optical path length variation coefficient of the fiber is $\sim 1.1\times 10^{-5}/°C$ [35], the temperature fluctuation of 2 °C will lead to an optical path length variation of $\sim 0.825$ m, corresponding to $\delta\tau_0 \approx 2.75 ns$. Therefore, this error term is very close to the result of zero-delay self-heterodyne interference, which is much smaller than the effective result $\varphi(t)-\varphi(t-\tau_0)$, and can be neglected in the following analysis.

For the second issue, it reflects environmental variations and persists even for the laser with a stable frequency. It can be expressed as:

$$\omega_c\delta\tau_0 = 2\pi f_c \frac{n_c\delta l}{c} = 2\pi \frac{n_c\delta l}{\lambda_c}. \quad (5)$$

Here, $\delta l$ is the fluctuation of fiber length. $\lambda_c$ represents the laser wavelength in vacuum, and $n_c$ is the refractive index of the optical fiber. When the environment changes, the impact of this error term is non-negligible. Considering that the range of optical path length $n_c\delta l$ is $\sim 0.825$ m, with $\lambda_c = 1550 nm$, the range of this error term can reach $\sim 10^6\pi$ rad. Therefore, this term is the primary error affecting the measurement.

To suppress the second error term and ensure the effectiveness of the digital decoherence method, the system is implemented as shown in Fig. 3(a). Laser T is LUT, and laser A is an auxiliary laser. Unlike the traditional heterodyne method, this system includes two interference systems corresponding to Laser T and Laser A, respectively. Both systems share the same delay fiber and AOM, which constitute the "common part". Since Laser A does not interfere directly with Laser T, it does not need to have the closely matched center wavelength with Laser T [36]. In detail, the beams from Laser T and Laser A are combined and transmitted through the common part. The delayed beams are then split and interfere with their respective reference beams from the two lasers. Band-pass filters, centered at the AOM's frequency, are used. As a result, in the "Interference system of Laser T", the beat signal between the reference beam and the delayed beam from Laser T is extracted, while other components are filtered out. Similarly, in the "Interference system of Laser A", the beat signal between the reference beam and the delayed beam from Laser A is obtained. Since both Laser T and Laser A propagate through the common part, the noise introduced by the common part is common-mode. The measurement result of Laser A can be used to compensate that of Laser T, which realizes the common-mode noise suppression for the second error term of Eq. (4). Furthermore, as the noise introduced by the common part, the phase noise of Laser A, and the phase noise of Laser T are mutually uncorrelated, the noise from the common part and Laser A can be suppressed based on the principle of correlation [37, 38], and the phase or frequency noise PSD of Laser T can then be extracted (a detailed discussion is provided in Sec. S1 in the Supplementary Material).

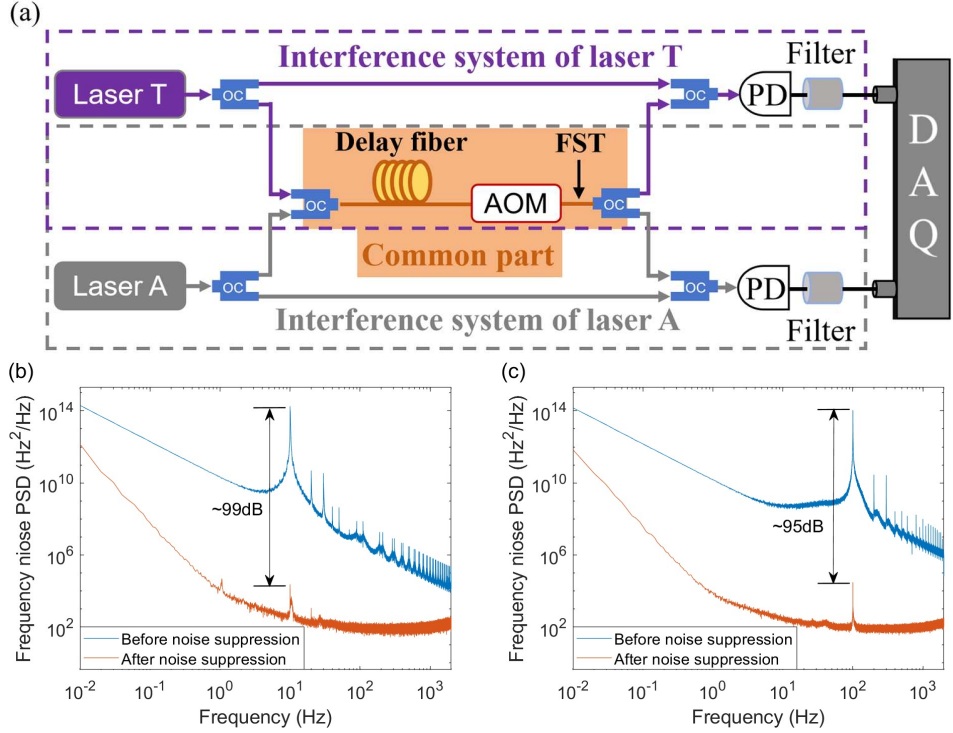

FIG. 3 (a) Experimental setup of the correlation system. OC: optical coupler; AOM: acousto-optical modulator; PD: photodetector; DAQ: data acquisition; FST: fiber stretcher. (b) The measurement system's capability to suppress fiber link noise at 10 Hz. (c) The measurement system's capability to suppress fiber link noise at 100 Hz.

To quantitatively evaluate the fiber noise suppression capability, a fiber stretcher (FST) is introduced into the "common part" of the system in Fig. 3(a) to apply external disturbances. First, a 10 Hz vibration is applied using the FST, and the frequency noise PSD measured by the "Interference system of Laser T" is shown as the blue curve in Fig. 3(b). A prominent 10 Hz component, along with its higher-order harmonics, is clearly observed in the PSD. In addition, the strong noise significantly affects the measured PSD at frequencies below 10 Hz. After applying common-mode noise suppression operation, the resulting frequency noise PSD is shown as the red curve in Fig. 3(b). The 10 Hz noise is reduced by ~99 dB. Conceivably, if we can generate a broad band noise (0.1Hz-1kHz), and raise the fiber noise by ~ 95 dB, the noise suppression ability of the common-mode noise suppression operation can be clearly quantified. However, it requires huge energy and cannot be realized by an FST. Consequently, we apply vibrations with the discrete frequency of 0.1Hz, 1Hz, 100Hz and 1kHz, respectively. Fig. 3(c) shows the noise suppression results when we apply a 100 Hz vibration, and ~95 dB noise suppression is achieved. In Supplementary Material S2, we show the noise suppression results at the frequency of 0.1Hz, 1Hz, and 1kHz, respectively. The measurement results reflect that the fiber noise can be suppressed by ~ 95 dB.

## 4. Results

### 4.1 Comparison with TCH method

To demonstrate the effectiveness of the digital decoherence method, we construct the experimental system shown in Fig. 3. The LUT, labeled as $T_1$, is an NKT Koheras BASIK X15 module with a typical linewidth of less than 100 Hz. The full noise PSD of laser with such narrow-linewidth is difficult to measure using the traditional DSHI method [26, 27]. The auxiliary laser, labeled A, is an Orion module from Redfern Integrated Optics company with ~kHz typical linewidth. These two lasers operate at different center wavelengths. Laser $T_1$ operates at 1550.12 nm, and Laser A operates at 1559.71 nm. A ~25 km fiber is used as the delay fiber, and the delay is measured to be $\tau_0 = 120267$ ns. Normally, the time delay can be measured via different methods [39] with ns-level accuracy. The frequency shift of the AOM is set to 100 MHz. Two band-pass filters, each centered at 100 MHz, are used to select the desired beating signals. A data acquisition (DAQ) system with a sampling rate of 120 MHz collects the beating signals from both systems, and its clock is locked to an atomic clock.

After 17 digital decoherence iterations (m=17), the equivalent decoherence time is increased to ~16 s. The resulting frequency noise PSD of Laser $T_1$ is shown as the red curve in Fig. 4(a). Simultaneously, a TCH system is set up using Laser $T_1$ and two other NKT Koheras BASIK X15 lasers, $T_2$ and $T_3$. The frequency noise PSD of Laser $T_1$ obtained using the TCH method is shown as the blue curve in Fig. 4(a). It can be seen that the result from the digital decoherence method is consistent with that obtained from the TCH method.

Furthermore, the linewidth of Laser $T_1$ is evaluated based on the frequency noise PSD and the β-separation line $\left[ (8\ln 2)/\pi^2 \right] f$ [40, 41] (a detailed discussion is provided in Sec. S3 in the Supplementary Material). The linewidth at different observation times is calculated and shown as the dash line in Fig. 4(a). The dash blue line represents the result obtained using the TCH method. At an observation time of 25 ms, it indicates a β-line linewidth of 53 Hz. Similarly, the dash red line, representing the result obtained using the digital decoherence method, shows a β-line linewidth of 47 Hz at the same observation time.

The frequency noise PSDs and linewidths of Lasers $T_2$ and $T_3$, measured under the same conditions, are shown in Figs. 4(b) and 4(c), respectively. When using the TCH method, all three lasers are uniformly set to a wavelength of 1550.12 nm. In contrast, when using the digital decoherence method, the three lasers operate at 1550.12 nm, 1550.04 nm, and 1550.20 nm, respectively, demonstrating the method's ability to measure lasers with different wavelengths. The frequency noise PSDs obtained using both methods are consistent. At an observation time of 25 ms, the digital decoherence method indicates a β-line linewidth of ~157 Hz for Laser $T_2$, and the TCH method obtains a β-line linewidth of ~162 Hz. For Laser $T_3$, the digital decoherence method shows a β-line linewidth of ~76 Hz, closely matching the β-line linewidth of ~77 Hz obtained using the TCH method.

Fig. 4(d) shows the frequency noise PSD and linewidth of the auxiliary laser A. The TCH method requires two additional lasers with the wavelength of ~1559.7 nm, which cannot be realized using the Laser T1, T2 and T3. Therefore, the measurement of the auxiliary laser A is conducted exclusively using the digital decoherence method. Compared to the NKT lasers, Laser A exhibits a performance gap of more than one order of magnitude, with a linewidth of ~5262 Hz measured at an observation time of 25 ms. Nevertheless, it can still serve as a valid auxiliary laser for effectively evaluating the NKT lasers when using the digital decoherence method.

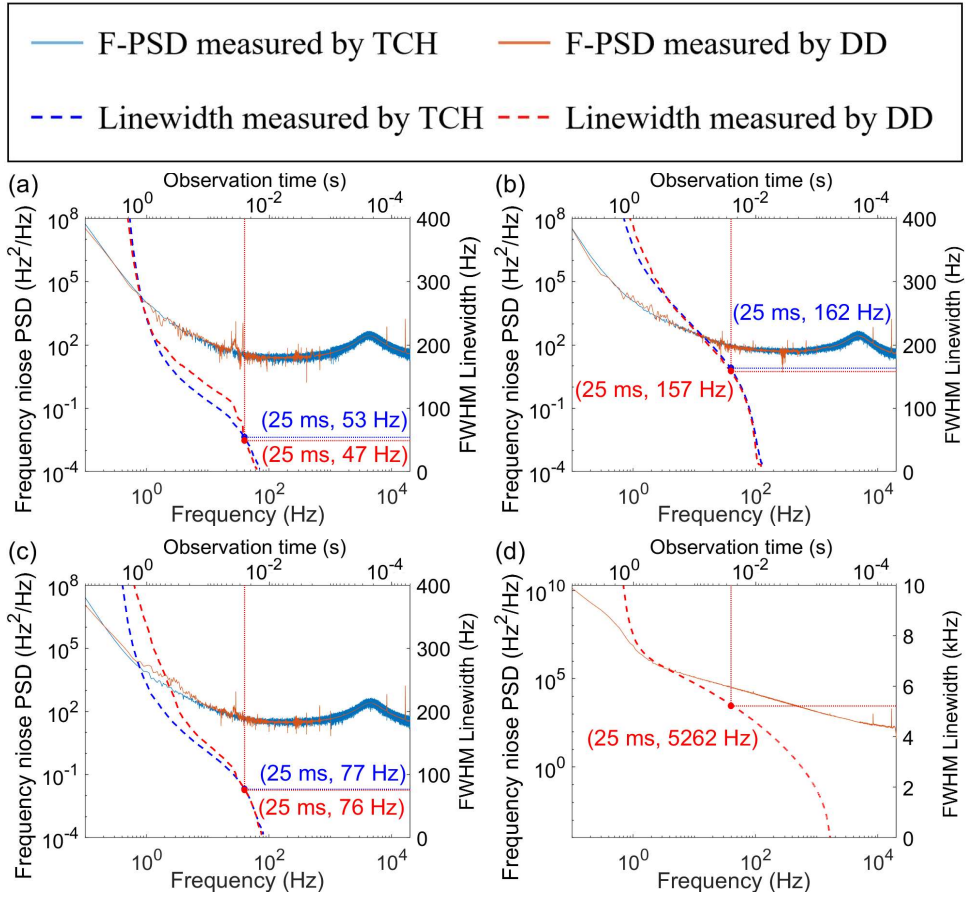

FIG. 4 The obtained frequency noise PSDs and linewidths. (a), (b), (c) The obtained frequency noise PSDs and linewidths of lasers T1, T2 and T3 using both the TCH method (blue line) and the digital decoherence method (red line). (d) The frequency noise PSD and linewidths of the auxiliary laser A obtained using the digital decoherence method. F-PSD: frequency noise PSD. DD: digital decoherence method.

### 4.2 Measurement result of sub-Hz laser

To demonstrate the measurement capability of the proposed method, two sub-Hz linewidth lasers from the optical clock group of NIM are used. One laser has a center wavelength of ~1542.4 nm. And the other laser has a center wavelength of ~1550.14 nm. The significant difference in their center wavelengths makes direct comparison impossible. Using the digital decoherence method and the system shown in Fig. 3(a), we can evaluate these two sub-Hz linewidth lasers. After 20 digital decoherence iterations (m=20), the equivalent decoherence time is increased to ~126 s, the relative frequency noise PSD between these two lasers is obtained, as shown by the red curve in Fig. 5(a). At an observation time of 10 s, the result indicates a β-line linewidth of 570 mHz. Besides, another method is used to calculate the minimum observable linewidth [42], which is 133 mHz (a detailed discussion is provided in Sec. S3 in the Supplementary Material).

Furthermore, we analyze the noises from various components of the system when the iteration times is 20. First, the noise induced by the delay fiber is measured (a detailed discussion is provided in Sec. S1 of the Supplementary Material), and the result is shown as the gray curve in Fig. 5(b). As the fiber-induced noise can be suppressed by up to ~95 dB using the common-mode noise suppression operation, the theoretical fiber noise floor can be evaluated and shown as the blue curve in Fig. 5(b).

In addition, we set the delay link length to zero, such that the measurement result excludes both laser phase noise and fiber-induced noise, leaving only the noise from the detection and acquisition system. The measurement result is shown as the red curve in Fig. 5(b). Based on the blue and red curves, the noise floor of the measurement system is obtained, which is shown as the black curve in Fig. 5(b). At an observation time of 25 s, the noise floor indicates a β-line linewidth of 210 mHz. The minimum observable linewidth of the noise floor is calculated to be 39 mHz (a detailed discussion is provided in Sec. S3 of the Supplementary Material).

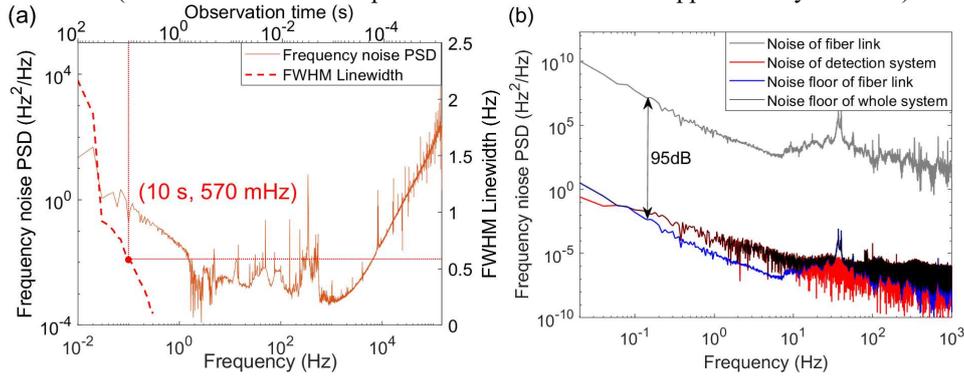

FIG. 5 (a) The frequency noise PSD and linewidths of two ultra-stable lasers' comparison result.
(b) Noise floor analysis. Gray: measured fiber link noise, red: detection system noise (zero delay),
blue: theoretical fiber link noise floor, black: noise floor of whole system.

## 5. Conclusion

In conclusion, we propose the concept of digital decoherence. It allows the decoherence process to be realized not only through physical optical path delay but also by digital delay, which simplifies the realization of decoherence, especially for long-coherence systems. It enables a compact laser spectrum analyzer which can measure the frequency noise spectra of sub-Hz lasers. It has broad wavelength compatibility and can simplify traditional measurements that rely on optical frequency combs. This analyzer provides an easy way to evaluate the laser quality in the fields of precision measurement, laser technology, and frequency standards.

**Funding.** National Natural Science Foundation of China (62171249).

**Disclosures.** The authors declare no conflicts of interest.

**Data availability.** Data underlying the results presented in this paper are not publicly available at this time but may be obtained from the authors upon reasonable request.